\documentclass[prl,twocolumn,preprintnumbers,amsmath,amssymb,superscriptaddress]
{revtex4}

\usepackage[pdftex]{graphicx, color}
\usepackage{amsmath}
\usepackage{amssymb}
\usepackage{amsthm}
\usepackage{amsfonts}						%
\usepackage{graphicx}
\usepackage{SIunits}
\usepackage[english]{babel}             	
\usepackage{graphicx}                   	
\usepackage{color}

\begin{document}

\title{Extraction of optical Bloch modes in a photonic-crystal waveguide}

\author{S.R. Huisman}\email{s.r.huisman@utwente.nl}
\affiliation{MESA+ Institute for
Nanotechnology, University of Twente, PO Box 217, 7500 AE Enschede, The Netherlands}
\author{G. Ctistis}
\affiliation{MESA+ Institute for
Nanotechnology, University of Twente, PO Box 217, 7500 AE Enschede, The Netherlands}
\author{S. Stobbe}
\affiliation{Niels Bohr Institute, University of Copenhagen, Blegdamsvej 17, DK-2100, Copenhagen, Denmark}
\author{J.L. Herek}
\affiliation{MESA+ Institute for
Nanotechnology, University of Twente, PO Box 217, 7500 AE Enschede, The Netherlands}
\author{P. Lodahl}
\affiliation{Niels Bohr Institute, University of Copenhagen, Blegdamsvej 17, DK-2100, Copenhagen, Denmark}
\author{W.L. Vos}
\affiliation{MESA+ Institute for
Nanotechnology, University of Twente, PO Box 217, 7500 AE Enschede, The Netherlands}
\author{P.W.H. Pinkse} \email{p.w.h.pinkse@utwente.nl, www.utwente.nl/mesaplus/anp}
\affiliation{MESA+ Institute for
Nanotechnology, University of Twente, PO Box 217, 7500 AE Enschede, The Netherlands}

\date{\today}

\begin{abstract}

We perform phase-sensitive near-field scanning optical microscopy on photonic-crystal waveguides. The observed intricate field patterns are analyzed by spatial Fourier transformations, revealing several guided TE- and TM-like modes. Using the reconstruction algorithm proposed by Ha, \textit{et al.} (Opt. Lett. 34 (2009)), we decompose the measured two-dimensional field pattern in a superposition of propagating Bloch modes. This opens new possibilities to study specific modes in near-field measurements. We apply the method to study the transverse behavior of a guided TE-like mode, where the mode extends deeper in the surrounding photonic crystal when the band edge is approached.

\end{abstract}
\maketitle

\section{Introduction}
Near-field scanning optical microscopy (NSOM) is a powerful tool to study objects with a resolution below the diffraction limit \cite{Novotny2006}. A unique feature of NSOM is the ability to tap light from structures that are designed to confine light, such as integrated optical waveguides \cite{Balistreri2001, Gersen2005} and cavities \cite{Louvion2006, Mujumdar2007, Lalouat2008, Lalouat2011}. Using the effect of frustrated total internal reflection, light that is invisible to other microscopy techniques can be detected. It is mainly for this reason that NSOM is so useful in the study of photonic-crystal waveguides. Photonic-crystal waveguides are two-dimensional (2D) photonic-crystal slabs with a line defect wherein light is guided \cite{Joannopoulos2008}. They possess unique dispersion relations, supporting slow-light propagation and enhanced light-matter interactions \cite{Krauss2007, Lund-Hansen2008}.

With NSOM, one can measure the dispersion relation in these waveguides, map light pulses spatially and study slow-light propagation \cite{Gersen2005, Engelen2005, Volkov2005}. It is also possible to measure the field patterns, which can be complicated because of the multimodal nature of the structures. Spatial Fourier transforms are especially useful to analyze \textit{e.g.}, the dispersion \cite{Gersen2005} or individual mode contributions \cite{Burresi2009}. For ballistic light propagation in photonic-crystal waveguides the detected field pattern is a superposition of Bloch modes determined by the symmetry of the waveguide. Ha \textit{et al.} \cite{Ha2009} recently proposed an algorithm that uses these symmetry conditions to extract Bloch modes from arbitrary measured field patterns \cite{Ha2011}. In the optical domain the algorithm has so far been used to identify dispersion relations in photonic-crystal waveguides \cite{Spasenovic2011, Ha2011-2}. To date, however, 2D spatial patterns of Bloch modes at optical frequencies have not been obtained with this method.

Here, we show the power of the Bloch mode reconstruction algorithm by extracting individual 2D mode patterns from phase-sensitive NSOM measurements on a GaAs photonic-crystal waveguide. We discuss in detail a specifically measured field pattern for which the Bloch modes are reconstructed. We apply this algorithm to study the behavior of the spatial width of the lowest frequency TE-like guided mode in the 2D band gap for TE-polarized light as a function of the wavevector.

\begin{figure}[]
  \includegraphics[width=8.5 cm]{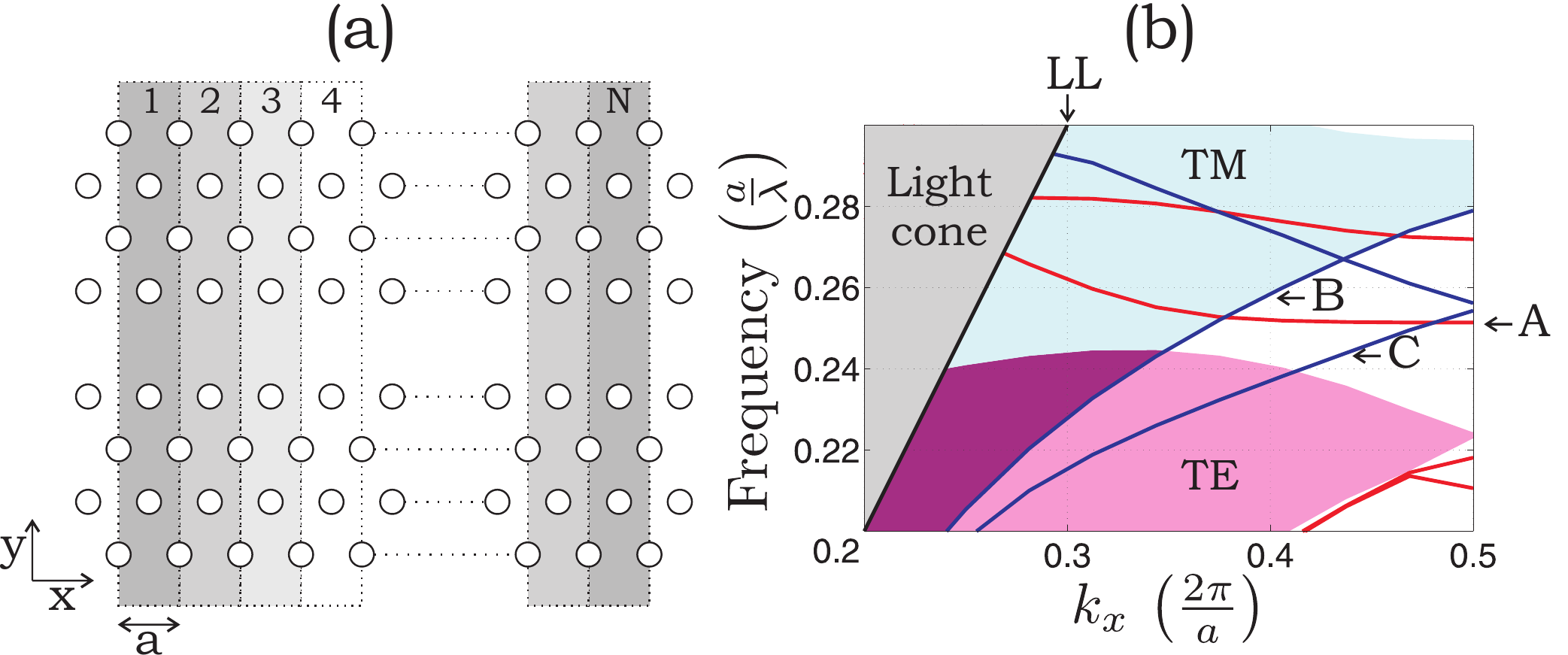}
\caption{\textit{(color online)} $(a)$ Top view of a photonic-crystal waveguide that is divided in $N$ unit cells (rectangles) for the reconstruction algorithm. $(b)$ Calculated bandstructure for a GaAs photonic-crystal waveguide showing both TE-like (red) and TM-like (blue) guided modes. The gray area marks the light cone, LL is the light line, the blue (pink) area marks TM-(TE)-like slab modes, the purple area marks both TE- and TM-like slab modes. The labeled arrows mark the different modes considered here.}
\label{fig1}
\end{figure}

\section{Samples and methods}

Figure \ref{fig1}$(a)$ illustrates the top view of the waveguide studied here. It consists of a GaAs photonic-crystal slab with holes forming a triangular lattice with pitch size $a=240 \pm 10$ nm, normalized hole radius $\frac{r}{a} = 0.309 \pm 0.002$, waveguide length of approximately 1 mm and slab thickness $h=160 \pm 10$ nm. A single row of missing holes forms a W1 waveguide. Light is guided in the $\hat{x}$-direction. Each numbered rectangle represents a unit cell. Fig. \ref{fig1}$(b)$ shows the calculated bandstructure along $\hat{x}$ for the photonic-crystal waveguide surrounded by air using a plane wave expansion \cite{mpb} (assuming a constant refractive index of $n_{\rm{GaAs}}=3.56$ and thickness $\frac{h}{a}=0.67$). The black diagonal line represents the light line, corresponding to light propagation in air. Modes above the light line couple to modes outside the waveguide, are therefore lossy \cite{Joannopoulos2008} and are not considered here. The blue and pink areas mark a continuum of modes propagating in the surrounding photonic crystal for TM- and TE-polarized light, respectively. The blue and red bands describe modes for, respectively, TM- and TE-polarized light that are guided by the line defect. Here we concentrate on mode $A$ that is TE-polarized, and modes $B$ and $C$ that are TM-polarized.

A continuous-wave diode laser (Toptica DL pro 940) is used with a linewidth of $100\,$kHz and an emission wavelength between $907-990\,$nm, corresponding to a reduced frequency in the range $0.24\ldots0.26\, a/\lambda$. Light is side-coupled on a cleaved end-facet of the 1 mm long waveguide with a high-NA glass objective (NA=0.55). The incident light is linearly polarized with an angle of about $45^o$ with respect to the normal of the waveguide to excite both TE- and TM-like modes. The field pattern is collected approximately $100$ \micro m away from the coupling facet using an aluminum coated fiber tip with an aperture of $160 \pm 10$ nm. We perform phase-sensitive NSOM using heterodyne detection. Detailed descriptions of a similar setup are presented in Ref. \onlinecite{Engelen2005}.

The propagating modes considered here represent eigenmodes of the crystal and can therefore be represented by Bloch modes. A 2D Bloch mode propagating in the ${\hat{x}}$-direction at position $\textbf{r}=[x,y]$ and frequency $\omega$ is described by $\Psi_m(\textbf{r},\omega) = \psi_m(\textbf{r}, \omega)\exp \left( ik_m \frac{x}{a} \right)$. Here $\psi_m(\textbf{r}, \omega)$ is an envelope that is periodic with the lattice and satisfies $\psi_m(\textbf{r}, \omega)=\psi_m(\textbf{r}+a\hat{x}, \omega)$, $m$ labels the Bloch mode, and $k_m$ is the corresponding normalized Bloch vector (we consider normalized wavevectors only). We restrict ourselves to propagating modes $\left( k \in \mathbb{R} \right)$. It is assumed that the measured field pattern $\Phi(\textbf{r}, \omega)$ can be described by a superposition of $M$ Bloch-modes $\Psi_m(\textbf{r}, \omega)$ with amplitude $a_m$ and one overall residual $\varepsilon(\textbf{r}, \omega)$:

\begin{equation}
\Phi(\textbf{r}, \omega)=\sum\limits_{m=1}^M a_m \Psi_m(\textbf{r}, \omega)+\varepsilon(\textbf{r}, \omega)
\label{eq1}
\end{equation}
The residual $\varepsilon(\textbf{r}, \omega)$ describes measured field patterns that cannot be described by the $M$ Bloch modes, such as non-guided modes, but also accounts for experimental artifacts and noise.

In the reconstruction algorithm \cite{Ha2009, Ha2011-2} a section of the waveguide is separated into $N$ unit cells, see Fig. \ref{fig1}$(a)$. The algorithm uses the property that each $\psi_m(\textbf{r}, \omega)$ is periodic in the photonic-crystal lattice, and requires us to analyze the measured field for each unit cell $U_n(\textbf{r'}, \omega)$, with $\textbf{r'}$ the coordinate within one unit cell. The measured $\Phi(\textbf{r},\omega)$ can be fitted with a series of Bloch modes using a least squares optimization that minimizes the functional $W=\int \vert \varepsilon (\textbf{r},\omega) \vert ^2 d\textbf{r} / \int \vert \Phi(\textbf{r},\omega) \vert ^2 d\textbf{r}$. This procedure results in the field pattern $A_m=a_m \psi_m(\textbf{r}, \omega)$ and the Bloch vector $k_m$ for $m=1\ldots M$. We compare the extracted $k_m$ with wavevectors determined from spatial Fourier transforms to confirm the accuracy of the algorithm.

\begin{figure}[]
  \includegraphics[width=8.5 cm]{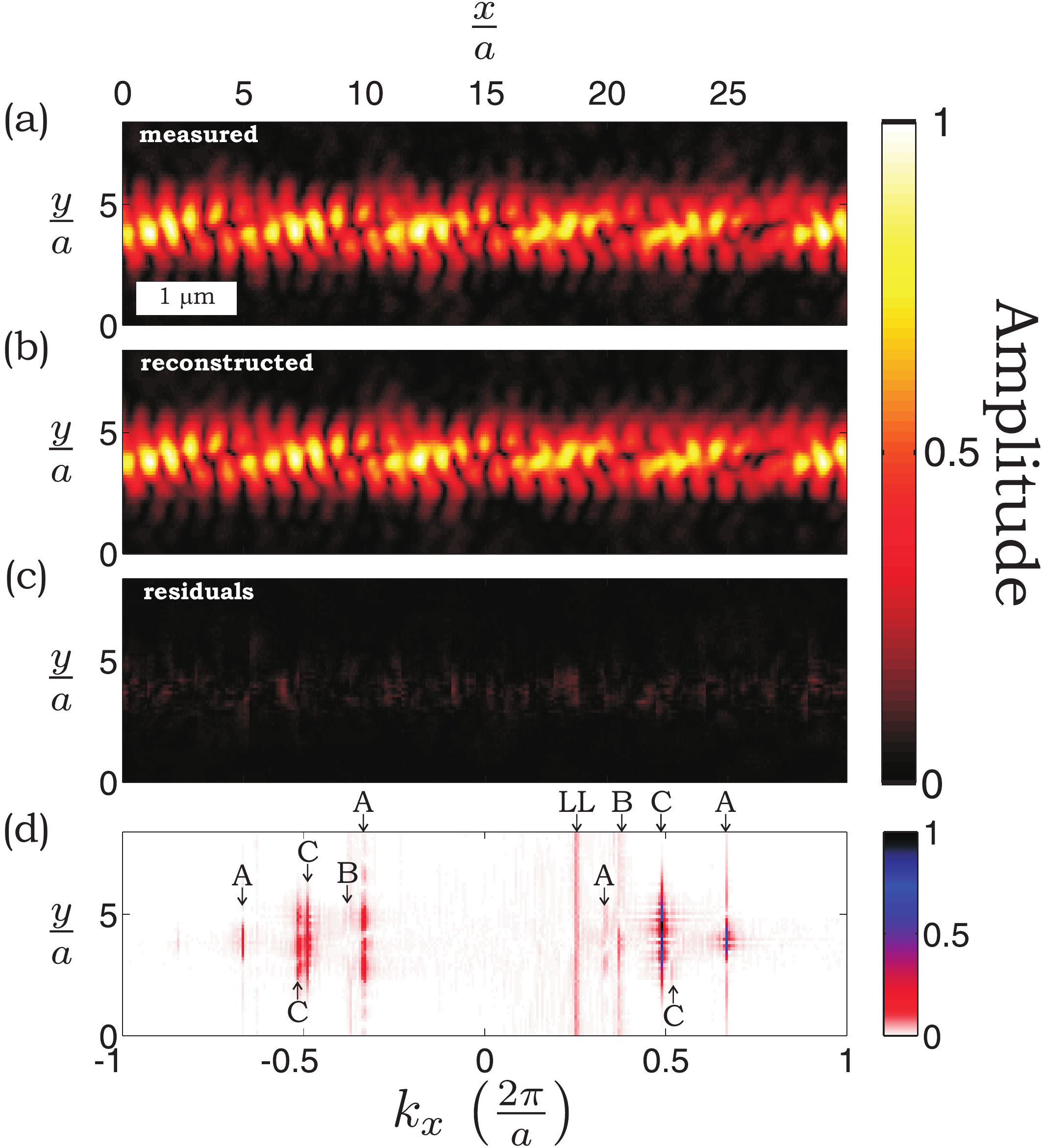}
\caption{\textit{(color online)} $(a)$ Measured amplitude for a photonic-crystal waveguide at $931.6 \pm 0.1$ nm. $(b)$ Fitted amplitude using 7 reconstructed Bloch modes over 30 unit cells. $(c)$ Residual amplitude. $(d)$ Amplitude coefficients of the spatial Fourier transforms of the measured near-field pattern. Labels correspond to those in Fig. \ref{fig1}$(b)$.}
\label{fig2}
\end{figure}
\section{Results}

Figure \ref{fig2}$(a)$ shows the measured field amplitude $\vert \Phi(\textbf{r}, \omega) \vert$ for a waveguide section of 30 unit cells ($7.2 \pm 0.5$ \micro m) at a wavelength of $\lambda = 931.6 \pm 0.1\,$nm. A beating pattern with a period of $5.7  \pm 0.2$ unit cells reveals that multiple modes are involved in the spatial pattern. To ensure that each $U_n(\textbf{r'}, \omega)$ describes precisely one unit cell, the original measurement was resampled on a different grid. 
Figure \ref{fig2}$(b)$ shows the reconstructed amplitude $\vert \sum\limits_{m=1}^M A_m (\textbf{r}, \omega) \vert$ with $M=7$ Bloch modes, which is in excellent agreement with the measured amplitude. We have chosen $M=7$ because we expect from Fig. \ref{fig1}$(b)$ a forward propagating mode corresponding to the light line and for modes A, B and C both forward and backward propagation, hence $M=1+(2 \times 3) =7$. The low values for the functional $W=0.0133$ demonstrates that indeed $\Phi(\textbf{r}, \omega)$ is well described by the superposition of 7 Bloch modes. This conclusion is confirmed by the absolute residuals $\vert \varepsilon(r, \omega) \vert$ plotted in Fig. \ref{fig2}$(c)$. The fitted $k_m$ are presented in the second column of Tab. \ref{tab1}. The third column describes the relative contribution of each mode as $c_m =\vert \int A_m^\ast (\textbf{r}, \omega) \Phi(\textbf{r}, \omega)  d\textbf{r} \vert/\int \vert \Phi(\textbf{r}, \omega) \vert ^2 d\textbf{r}$. Note that the 7 contributions plus that of the residual add up to unity. The fifth column describes which modes of Fig. \ref{fig1}$(b)$ correspond to $k_m$, the propagation direction and polarization. We observe mainly the forward propagating TE-like mode $A$ ($m=5$) and the forward propagating TM-like mode $C$ ($m=4$). The errors in $k_m$ are estimated by varying the grid element size and allowing for a relative increase of $\Delta W$ by maximum $10\%$; within this range the mode patterns $A_4 (\textbf{r}, \omega)$ and $A_5 (\textbf{r}, \omega)$ do not change noticeably.

Next, the fitted $k_m$ are compared with wavevectors determined from the spatial Fourier transforms shown in Fig. \ref{fig2}$(d)$ $(k_{\rm{SFT}})$. A Fourier transform in the $\hat{x}$-direction was made for each line parallel to the waveguide over a range of $35.4 \pm 0.9$ \micro m, which includes the range shown in Fig. \ref{fig2}$(a)$. For comparison $k_m$ and $k_{\rm{SFT}}$ are listed in Tab. \ref{tab1}, showing an excellent agreement. The spatial Fourier transforms show for modes $A$ and $C$ higher Bloch harmonics. Both the fundamental $k_{\rm{SFT}}$ and the observed higher Bloch harmonics are listed in the table. In Fig. \ref{fig2}$(d)$ the modes from Fig. \ref{fig1}$(b)$ are identified. The amplitude coefficients confirm that we detect mainly the forward propagating TE-like mode $A$ ($k_{\rm{SFT}}=-0.332 \pm 0.003, 0.668 \pm 0.003$) and the forward propagating TM-like mode $C$ ($k_{\rm{SFT}}=-0.516 \pm 0.003, 0.489 \pm 0.003$).

\begin{table}

\caption{Comparison between fitted wavevectors from Bloch-mode reconstruction ($k_m$), and obtained wavevectors from spatial Fourier transforms ($k_{\rm{SFT}}$). The first column labels the Bloch modes. The second column gives the fitted $k_m$. The third column gives a measure how strongly present a mode is. The fourth column gives the $k_{\rm{SFT}}$, where the superscript $F$ marks the fundamental wavevector. In the fifth column we identify the modes from the calculated bandstructure in Fig. \ref{fig1}$(b)$, the propagation direction, where $+(-)$ corresponds to the positive(negative) $\hat{x}$-direction, and polarization.}
\begin{small}

\begin{tabular}{crccr}

	\hline
$m$	&	$k_m$ $(\frac{2 \pi}{a})$	&  $c_m $	& $k_{\rm{SFT}}$ $(\frac{2 \pi}{a})$		& Label  \\	\hline	\hline
$1$   & 	$0.251(1)$ 		& $0.0073$	&	$0.251(3)$ 							& +$LL$		\\
$2$   & 	$0.330(3)$		& $0.0117$	&	$-0.668(3)^F$, $0.331(3)$				& -$A$, TE	\\
$3$   & 	$0.379(2)$		& $0.0086$	&	$0.370(3)$							& +$B$, TM	\\
$4$   & 	$0.488(2)$		& $0.6714$	&	$-0.516(3)$, $0.489(3)^F$  			& +$C$, TM 	\\
$5$   & 	$-0.331(2)$		& $0.2202$	&	$-0.332(3)$, $0.668(3)^F$ 			& +$A$, TE	\\
$6$   & 	$-0.367(6)$  	& $0.0014$	&	$-0.370(3)$  						& -$B$, TM	\\
$7$   & 	$-0.486(2)$  	& $0.0681$	&	$-0.489(3)^F$, $0.516(3)$ 			& -$C$, TM	\\
	\hline
	\label{tab1}
\end{tabular}\end{small}
\end{table}
\begin{figure}[]
  \includegraphics[width=8.5 cm]{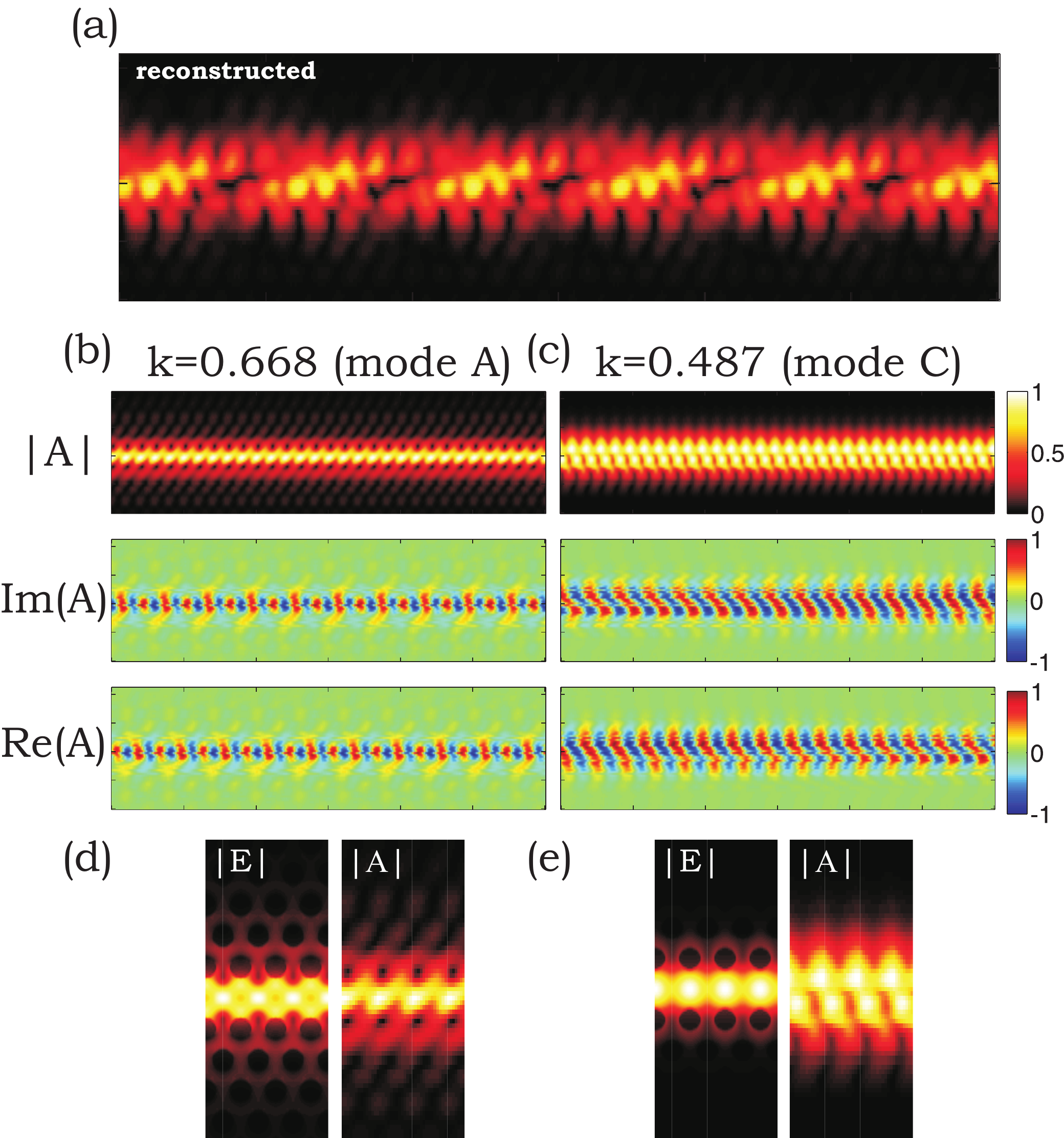}
\caption{\textit{(color online)} $(a)$ Reconstructed amplitude for Fig. \ref{fig2}$(a)$ using Bloch modes $m=4$ and $m=5$ only. $(b-c)$ Amplitude, imaginary part and real part for both Bloch modes. $(d-e)$ Comparison between calculated $|E|$ (left) and reconstructed $|A|$ (right). }
\label{fig3}
\end{figure}

We have demonstrated that the forward propagating TE-like mode $A$ and the forward propagating TM-like mode $C$ are the most prominent Bloch modes present in the data of Fig. \ref{fig2}. Figure \ref{fig3}$(a)$ shows the amplitude when only these two modes are taken into account for the reconstruction. A very good agreement is observed with $\Phi(\textbf{r}, \omega)$ of Fig. \ref{fig2}$(a)$. Especially the diagonal beats are well reproduced. The difference wavevector of the two modes corresponds to a beating period of $(5.5 \pm 0.2)a$. The beating pattern of two orthogonal modes is the result of quasi-interference; the NSOM tip thereby projects both orthogonal polarizations on a detection basis where these modes interfere \cite{Balistreri2000}. Figure \ref{fig3}$(b)$ shows the amplitude, the real part and the imaginary part of the reconstructed TE-like Bloch mode $A$ with $k_5=-0.331 \pm 0.002$. The mode profile is symmetric in the $\hat{y}$-direction about the center of the waveguide. Figure \ref{fig3}$(d)$ shows the calculated \cite{mpb} time-averaged amplitude $\langle|E|\rangle$ (left) and the measured amplitude $|A|$ for approximately 3 unit cells. Both show a similar pattern. Figure \ref{fig3}$(c)$ shows the amplitude, the real part and the imaginary part of the reconstructed TM-like Bloch mode $C$ with $k_4=0.487 \pm 0.003$. Figure \ref{fig3}$(e)$ shows the calculated $\langle|E|\rangle$ (left) and the measured amplitude $|A|$ for approximately 3 unit cells (right). For mode $C$ the agreement is poor, likely because the near-field tip has a low response to $E_z$ and a non-trivial response to $E_x$ and $E_y$, see Ref. \onlinecite{Burresi2009}, in addition to its finite resolution. The reconstructed field patterns are an approximation of the Bloch modes propagating in the system and moreover, they are not necessarily orthogonal because of quasi-interference. Although the effect of the tip is far from straightforward (\cite{Hopman2006, Mujumdar2007, Lalouat2007}), we anticipate that the comparison between calculated modes of an optical system and reconstructed modes could lead to methods to deduce the response function of a near-field tip.

\begin{figure}[]
  \includegraphics[width=8.5 cm]{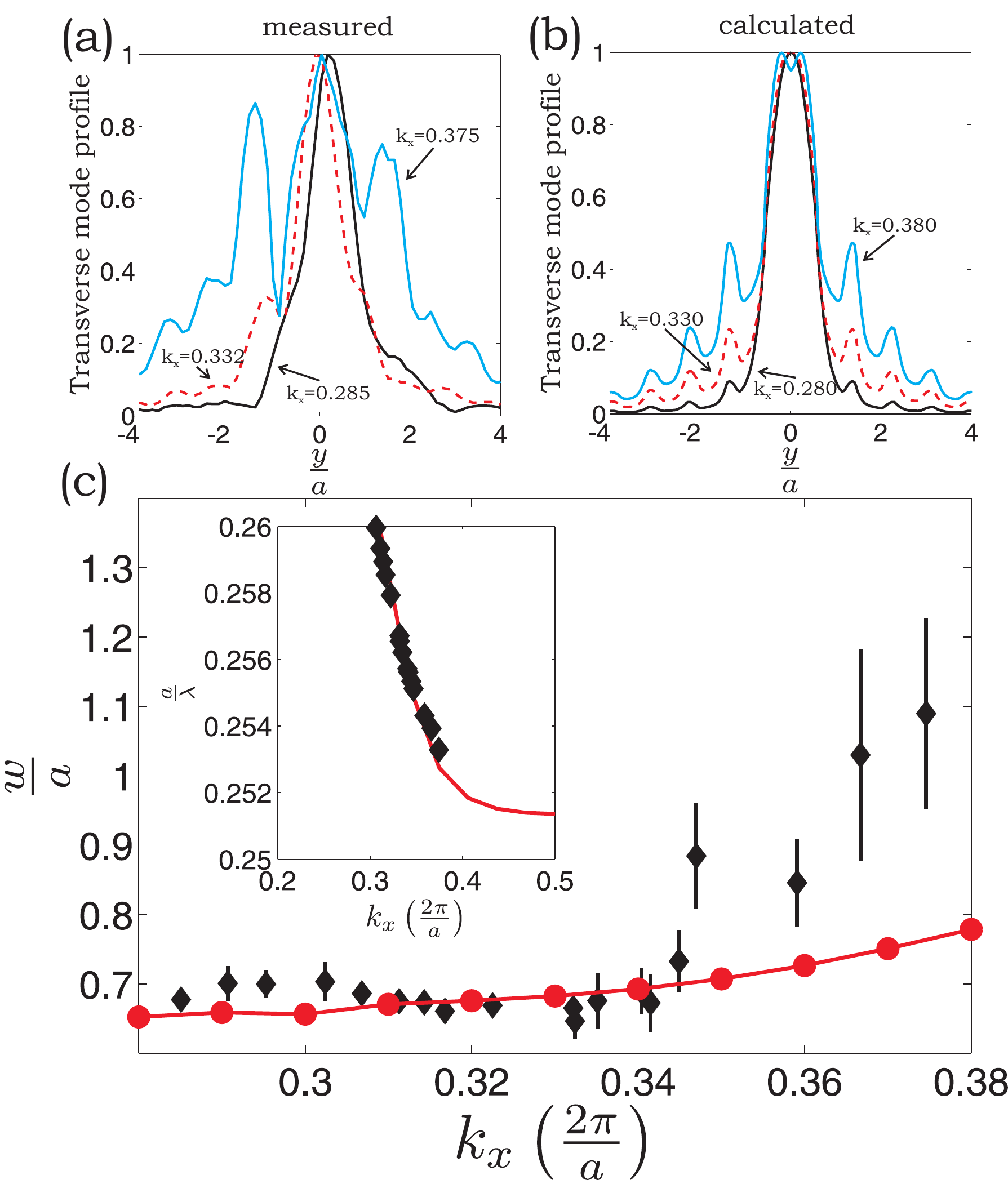}
\caption{\textit{(color online)} $(a)$ Measured normalized transverse mode profile  for $k_x=0.285$ (black), $k_x=0.332$ (red dashed), and $k_x=0.375$ (blue). $(b)$ Calculated normalized transverse mode profile for $k_x=0.280$ (black), $k_x=0.330$ (red dashed), and $k_x=0.380$ (blue). $(c)$ Determined (black) and calculated width $w$ (red) versus longitudinal wavevector $k_x$ for the TE-like guided mode. Inset: measured (symbols) and calculated dispersion (red). }
\label{fig4}\end{figure}

Next, we demonstrate the power of the reconstruction algorithm by studying the transverse behavior of TE-like mode (A) versus wavevector. We consider the forward propagating TE-like mode ($k_5=-0.331, 0.669$ in Fig. \ref{fig3}) and take its reduced wavevector ($0 < k_x < 0.5$) to compare directly with the folded bandstructure of Fig. \ref{fig1}$(b)$. We have measured field patterns between $\lambda=907-944$ nm, and apply the reconstruction algorithm to determine $A(\textbf{r}, \omega)$ and $k_x$ for this mode at each $\lambda$. We have selected $\Phi(\textbf{r}, \omega)$ where this TE-like mode is prominently present in spatial Fourier transforms. The inset in Fig. \ref{fig4}$(c)$ shows the fitted $k_x$ versus reduced frequency $\frac{a}{\lambda}$.

In order to concentrate on the transverse behavior, we define the transverse mode profile $\int\limits_{0}^{a} \vert A(\textbf{r}, \omega) \vert dx$. Figure \ref{fig4}$(a)$ shows the measured normalized transverse mode profile for 3 different wavevectors. At $k_x=0.285$ (black) a transverse mode profile is apparent that can be mainly described by one prominent maximum at $\frac{y}{a}=0$ that is slightly asymmetric, describing light guided in the line defect.  Additional side lobes are observed at $\frac{y}{a}=-0.9$ and at $\frac{y}{a}=1.4$, representing light extending into the surrounding photonic crystal. At $k_x=0.332$ (red dashed) we observe a central maximum at $\frac{y}{a}=0$, and the contributions of the side lobes become bigger. Also note the new peaks observed at $\frac{y}{a}=-3.2, -2.2, -1.5$, and $2.8$. When the wavevector is increased, the relative contributions of these additional peaks increase. At $k_x=0.375$ the central peak is still present, and the surrounding peaks have grown. Qualitatively, the measured transverse mode profiles correspond with the calculated ones shown in Fig. \ref{fig4}$(b)$, which were obtained from the time-averaged amplitude of the total electric field $\int\limits_{0}^{a} \vert E(\textbf{r}, \omega) \vert dx$. The maxima and minima occur at approximately the same locations and the width $w$ of the central maximum is growing with increasing $k_x$. Not all features are resolved of the calculated transverse mode profile in our measurements. For example, the measured relative amplitude of the central maximum compared with the additional maxima differs from the calculations.

The central maxima of the transverse mode profiles are fitted with a Gaussian with width $w$. Figure \ref{fig4}$(c)$ shows the reduced width $\frac{w}{a}$ versus wavevector (black symbols, bars represent $95\%$ convergence intervals). The red symbols interpolated by the dashed line represent $\frac{w}{a}$ determined from the calculated transverse mode profile. The measured $w$ increases with $k_x$. For $k_x<0.34$ the measured $w$ matches the calculated $w$ well. For $k_x>0.34$ the measured $w$ becomes larger than the calculated $w$. We attribute this to the finite resolution of the near-field tip and to its response function. For most considered $k_x$-values, however, the measured $w/a$ matches the calculated with a $10\,\%$ accuracy.

\section{Summary}

In conclusion, we have implemented an algorithm proposed by  Ha \textit{et al.} to extract Bloch modes from near-field measurements on a photonic-crystal waveguide. The extracted wavevectors are in very good agreement with the wavevectors determined from spatial Fourier transforms. We have studied two extracted Bloch modes to explain the observed near-field pattern. We also have studied how the width of a selected mode changes with wavevector and find good agreement with calculations. We anticipate that this algorithm can be used to filter states that cannot be described by propagating Bloch modes, such as Anderson-localized states observed in the slow-light regime \cite{Topolancik2007, Sapienza2010, Garcia2010, Yang2011, Huisman2011}.

\section*{Acknowledgments}
We kindly thank Dirk Jan Dikken, Herman Offerhaus, Kobus Kuipers and Allard Mosk for stimulating discussions, Cock Harteveld, Jeroen Korterik, and Frans Segerink for technical support. This work was supported by FOM and NWO-Nano.

\end{document}